\documentclass[preprintnumbers,twocolumn,amsmath,amssymb,dvips,floatfix,superscriptaddress,showpacs]{revtex4}
\usepackage{dcolumn}
\usepackage{graphicx} 

\begin{document}

\title{Surface Induced Crystallization in Supercooled Tetrahedral
Liquids}

\author{Tianshu Li} \email[corresponding author:]{tsli@ucdavis.edu}
\affiliation{Department of Chemistry, University of California, Davis,
CA 95616} \author{Davide Donadio} \affiliation{Department of
Chemistry, University of California, Davis, CA 95616} \author{Luca
M. Ghiringhelli} \affiliation{Max-Planck-Institute for Polymer
Research Theory Group, PO Box 3148 D-55021 Mainz, Germany}
\author{Giulia Galli} \affiliation{Department of Chemistry, University
of California, Davis, CA 95616}


\begin{abstract}
Freezing is a fundamental physical phenomenon that has been studied
over many decades; yet the role played by surfaces in determining
nucleation has remained elusive. Here we report direct computational
evidence of surface induced nucleation in supercooled systems with a
negative slope of their melting line ($dP/dT < 0$). This unexpected
result is related to the density decrease occurring upon
crystallization, and to surface tension facilitating the initial
nucleus formation. Our findings support the hypothesis of surface
induced crystallization of ice in the atmosphere, and provide insight,
at the atomistic level, into nucleation mechanisms of widely used
semiconductors.
\end{abstract}


\maketitle

In 1910, F. Lindemann \cite{Lindemann} suggested that melting of a
crystal begins when the root-mean-square amplitude of the lattice
vibration reaches a critical fraction of the nearest neighbor
distance. As surface atoms are usually weakly bonded and
under-coordinated compared to those in the bulk, melting is often
observed to originate at the surface \cite{Cahn} both in experiments
\cite{Frenken} and computer simulations \cite{Tartaglino}.  The
nucleation of crystals from the melt is in turn a more complex
phenomenon.  In the absence of heterogeneous centers, homogeneous
nucleation occurs, and the effect of surfaces on this process is not
well understood.  While common wisdom would regard surfaces as
unfavorable nucleation sites, atmospheric observations and
sophisticated experiments on suspended droplets of water
\cite{Tabazadeh,Shaw,Durant} and of liquid metal alloys
\cite{Sutter,Shpyrko} support the hypothesis of surface-induced
crystallization \cite{Djikaev} for some systems.

Despite its remarkable implications in a variety of scientific
disciplines, such as atmospheric physics \cite{Tabazadeh,Shaw,Sastry},
metallurgy \cite{Shpyrko}, and nanoscience \cite{Sutter}, a
microscopic understanding of nucleation process in the presence of
free surfaces is still missing. Technical difficulties in designing
experiments to capture nucleation events in, {\em e.g.}, suspended
droplets, have so far prevented an accurate characterization of the
freezing process and direct simulations of nucleation require very
challenging computer experiments, as the time scales involved usually
exceed the capabilities of present day computers.  In the recent
literature, nucleation rates were computed for simple systems by
employing accelerated Monte Carlo or molecular dynamics (MD)
algorithms \cite{Bennett,Chandler,vanDuijneveldt}, {\em e.g.}, in
Lennard-Jones liquid \cite{Moroni,Trudu}, hard-sphere colloids
\cite{Auer, Cacciuto}, and two-dimensional Ising model \cite{Sear}.

In this Letter, we combine the recently developed forward flux
sampling (FFS) method \cite{Allen1,Allen2} with MD simulations
\cite{Voter} to compute the nucleation rate of supercooled liquid Si,
at temperatures $T$ up to 95\% of the melting point.  Our study shows
that the presence of free surfaces may enhance the nucleation rates by
several orders of magnitude with respect to those found in the bulk,
and demonstrate that free surfaces, in addition to their well-known
role in initiating melting, can also be ``catalytic'' sites for
freezing in tetrahedral liquids with a negative slope of their melting
lines ($dP/dT<0$).

We carry out MD simulations, within the isobaric-isothermal canonical
ensembles (NPT) at $p=0$, and within the isothermal canonical
ensembles (NVT), for bulk Si and for slab configurations,
respectively. Most of our simulations are performed with 5832 atoms in
a cubic cell using Tersoff potential \cite{Tersoff}. The rate for
growing a solid cluster containing $\lambda$ atoms out of the liquid
can be expressed \cite{van-Erp} by the product of a flux rate
$\dot{\Phi}_{\lambda_0}$ for the formation of smaller clusters with
$\lambda_0$ ($\lambda_0<\lambda$) atoms, and the probability
$P(\lambda|\lambda_0)$ for these clusters to eventually grow to size
$\lambda$. Within the FFS scheme one can compute these two terms
separately, and $P(\lambda|\lambda_0)$ is obtained by sampling a
number of interfaces between the initial and final states in the space
of the order parameter. In our case it is natural to choose this
parameter as the number of Si atoms $\lambda$ formed in the largest
solid cluster. To identify Si crystalline clusters in the liquid, we
employ a local order parameter, $q_3$ as defined in
Ref. \cite{Ghiringhelli}, which has been shown to be highly sensitive
to crystalline order \cite{Steinhardt}. The flux rate
$\dot{\Phi}_{\lambda_0}$ is computed by conducting standard MD
simulations starting from a well-defined basin ($\lambda<\lambda_A$)
in phase space. As occasional fluctuations lead to direct crossing of
the first interface, $\lambda_0$, the atomic configuration is
recorded. The simulation is then continued until $N_0$ ($\sim 120$)
configurations are collected and the average flux rate is given by
$N_0/(t_0V)$, where $t_0$ and $V$ are the total simulation time and
the system volume, respectively. During this step, we find that
inclusion of free surfaces does not significantly change the flux
rates, at all temperatures. Therefore the influence of free surfaces
on nucleation rates is expected to be sufficiently well represented by
the variation of the growth probability $P(\lambda|\lambda_0)$.

\begin{figure}[t]
\begin{center}
\includegraphics[width=3.0in]{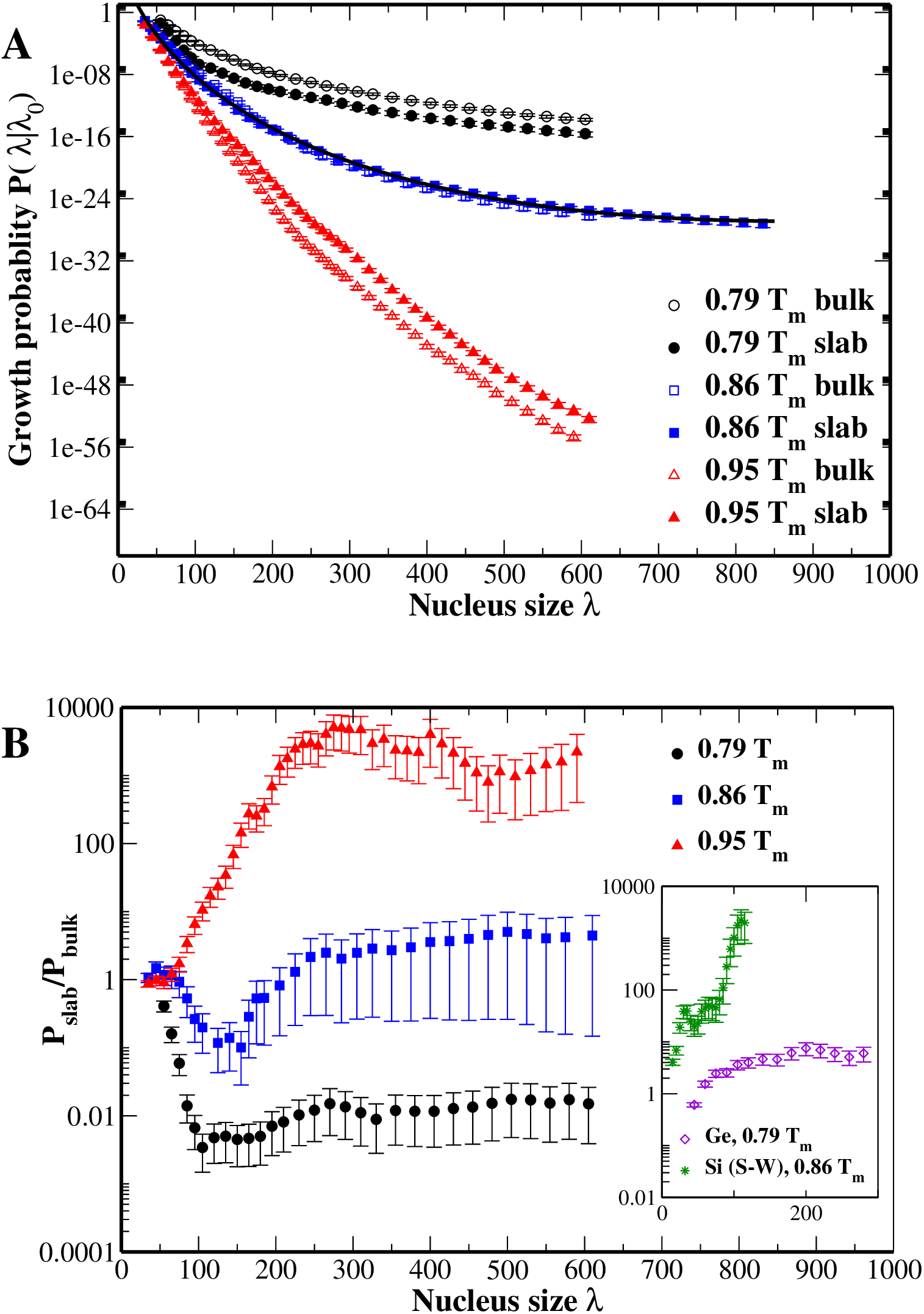}
\caption{(color online) (\textbf{A}) Calculated growth probability
  $P(\lambda|\lambda_0)$ as a function of the cluster size in both
  bulk liquid Si and liquid slab at different temperatures. At 0.86
  $T_m$ $P(\lambda|\lambda_0)$ is fitted by
  $A\,\mbox{exp}[B\lambda^{2/3}+C\lambda]$ (solid line), where A, B,
  and C are fitting constants. The nucleation rate at 0.86 $T_m$ (see
  text) is given by $\dot{\Phi}_{\lambda_0} \,P(\lambda_c|\lambda_0)$,
  where $\lambda_c=(-3C/2B)^{-3}$ is the extrapolated critical
  size. (\textbf{B}) Calculated ratio of the growth probability
  between the liquid slab and the bulk liquid,
  $P_{\mbox{slab}}/P_{\mbox{bulk}}$, as a function of the cluster
  size. The insert shows the same ratios for Tersoff Ge \cite{Tersoff}
  at 0.79 $T_m$ and Stillinger-Weber \cite{SW} Si at 0.86 $T_m$. The
  error bar is computed based on the binomial distribution of $k_i$,
  the number of successful trial runs at $\lambda_i$ \cite{Allen2}.}
\label{f1}
\end{center}
\end{figure}

To compute the growth probability $P(\lambda|\lambda_0)$, we start
from the configurations collected at the interface $\lambda_0$ and
carry out a large number ($M_1$, typically around 1000$\sim$10,000) of
trial MD runs with different randomized initial momenta. A few ($k_1$)
trial runs result in successful crossings to the next interface, while
in the remaining cases small crystalline clusters dissolve.  The
individual crossing probability $P(\lambda_1|\lambda_0)$ is then given
by $k_1/M_1$. The subsequent trial runs are launched at these crossing
points on the next interface and the total growth probability is given
by: $P_{\lambda_n}=\prod_{i=1}^n P(\lambda_i|\lambda_{i-1})$
\footnote{In order to evaluate the effect of sampling techniques on
our results, we repeat our calculations of $P(\lambda|\lambda_0)$ at
0.95 $T_m$ by employing a Langevin thermostat \cite{Chandrasekhar},
and by varying both the interface spacing and the cell size. The
observed changes in nucleation rate are within the error bars given in
Fig.1A.}.

The calculated transition probability $P(\lambda|\lambda_0)$ as a
function of the cluster size $\lambda$ is shown in Fig.1A for both
bulk systems and slabs, at several temperatures. Initially
$P(\lambda|\lambda_0)$ sharply decreases as small clusters grow, and
then it tends to saturate, indicating the formation of a critical size
nucleus. Consistent with classical nucleation theory, the calculated
nucleation rates show a strong dependence on $T$: Raising $T$ from
0.79 $T_m$ to 0.86 $T_m$ yields a significant decrease in the rate by
over 12 orders of magnitude. In particular, the computed nucleation
rate $1.14\pm0.89\times10^{10} \mbox{m}^{-3}\mbox{s}^{-1}$ at 0.86
$T_m$ agrees well with the experimental measurement
$2.0\times10^{10}\mbox{m}^{-3}\mbox{s}^{-1}$ at $14\pm1$\%
supercooling \cite{Devaud}.

The most striking result of this calculation is that it demonstrates a
clear transition in preferential nucleation from the bulk to the slab,
as $T$ increases. This finding is illustrated in Fig.1B that shows the
ratio of the growth probability between the liquid slab and bulk,
$P_{\mbox{slab}}/P_{\mbox{bulk}}$, as a function of $\lambda$. At 0.79
$T_m$, this ratio decreases rapidly, as the small clusters grow, and
it stabilizes around $10^{-2}$, for $\lambda>100$.  This indicates
that nucleation from the liquid slab is unfavorable as compared to
that from the bulk. When $T$ is raised to 0.86 $T_m$,
$P_{\mbox{slab}}/P_{\mbox{bulk}}$ is around unity, suggesting that in
the slab and the bulk one achieves virtually identical nucleation
rates. As the temperature is further increased up to 0.95 $T_m$, the
liquid slab yields a nucleation rate over a thousand times higher than
that in the bulk liquid, for $\lambda\sim 200$.

\begin{figure*}
\begin{center}
\includegraphics[width=5.0in]{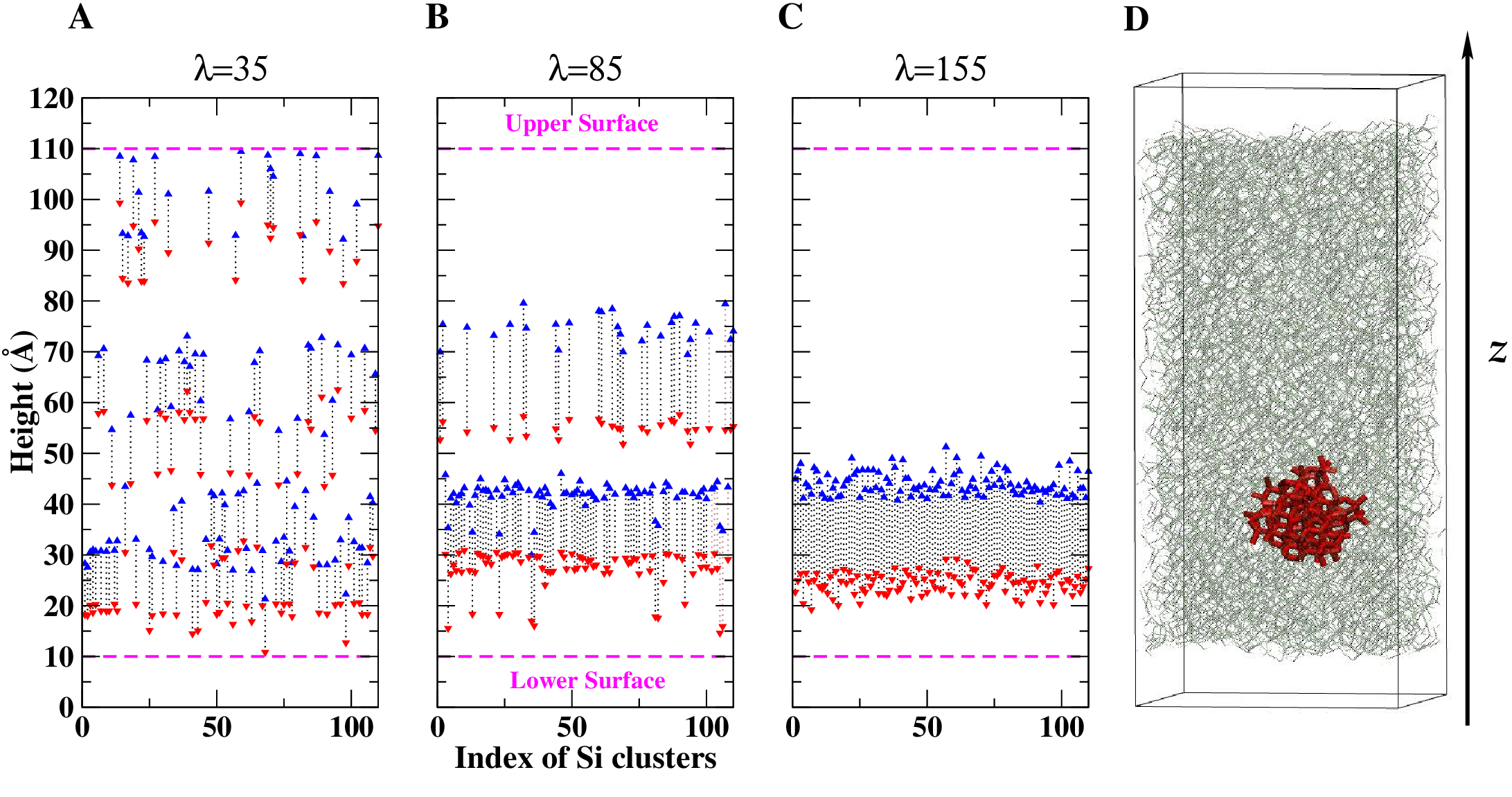}
\caption{(color online) Distributions of solid Si crystallites normal
  to the free surface in the Si liquid slab as a function of the
  cluster size (\textbf{A}) $\lambda=35$; (\textbf{B}) $\lambda=85$;
  and (\textbf{C}) $\lambda=155$ at 0.95 $T_m$. Each vertical pair of
  blue (red) triangles up (down) connected by a dashed line represent
  the upper (lower) boundaries of a solid Si cluster. Panel
  (\textbf{D}) shows a snapshot of a solid Si cluster (red) containing
  155 Si atoms (corresponding to one of the configurations in
  (\textbf{C}) surrounded by liquid Si (gray line) in the slab with
  two free surfaces, normal to the $z$ direction.}
\label{f2}
\end{center}
\end{figure*}
   
This noticeable increase in nucleation rates is then naturally
attributed to the presence of the free surface in the slab.  To show
this is indeed the case, we explore the microscopic details of growing
Si clusters, particularly their distributions in the direction $z$
normal to the free surface. Here we replicate the unit cell and
include 11,664 atoms. Fig.2 displays such distributions at different
stages of the cluster formation. Initially, the small solid Si
clusters are distributed nearly evenly along the $z$ axis (Fig.2A),
which is in accord with the computed flux rates
$\dot{\Phi}_{\lambda_0}$ being of comparable magnitude in the bulk and
in the slab. As solid clusters grow, there appears a clear tendency
for those with the highest growth probability to be located close to
the free surface. Such a tendency is confirmed by fewer clusters being
present in the middle of the slab, as $\lambda$ increases.  Finally
all clusters exclusively reside about 1 nm away from the immediate
interface between the liquid and vacuum (Fig.2C).

To understand the ``catalytic'' role of the liquid surface in
crystallization events, we consider the nucleation rate $R=A\,
\mbox{exp}(-\Delta G^{*}/k_BT)$, where $A$ is a kinetic pre-factor and
$\Delta G^{*}$ is the minimum free energy barrier.  Free surfaces do
not exhibit the same order as in the bulk, and have undercoordinated
atoms, making the formation of crystalline clusters unfavorable in
their vicinity, if the bulk offers a sufficient number of atomic sites
for a nucleus to grow. On the other hand, the generally high atomic
mobility near free surfaces contributes to an enhancement of the
kinetic factor $A$. The free energy change $\Delta G$ for the
formation of a small crystallite is the sum of the volume contribution
$\Delta G_{\mbox{v}}$ and the solid-liquid interface contribution
$\Delta G_{\mbox{i}}$: $\Delta G=\Delta G_{\mbox{v}}+\Delta
G_{\mbox{i}}$. Observing that in the liquid slab the solid Si clusters
reside in the subsurface (see Fig.2 C and D) and are thus still
surrounded by a liquid like environment, we assume that the
solid-liquid interface contribution $\Delta G_{\mbox{i}}$ remains the
same near the free surface; the volume contribution $\Delta
G_{\mbox{v}}$ is instead decreased, as compared to the bulk. In
particular in our simulations we find that the free surface introduces
a small lateral pressure field ($p<0$), in the plane parallel to the
surface.  Therefore a pressure dependent term must be added to the
volume free energy change $\Delta G_{\mbox{v}}$, in order to account
for the nucleation of a cluster containing $\lambda$ atoms: $\delta
G_{\mbox{v}}(p)=\lambda p(\rho_L-\rho_S)/(\rho_L\rho_S)$, where
$\rho_L$ and $\rho_S$ are the number densities of the liquid and
solid, respectively.  Since liquid Si is denser than the solid at the
melting point, {\em i.e.}, $\rho_L>\rho_S$, $\delta G_{\mbox{v}}(p)$
is negative. It then follows that the energy barrier for nucleation is
slightly lowered near a liquid surface, relative to that in the bulk
where $p=0$. In other words, as in a tetrahedral liquid with $dP/dT<0$
the density is decreased upon solidification, the presence of a free
surface can accommodate volume expansion more easily due to surface
tension, and thus nucleation in its vicinity may be preferred. To
further elucidate the role of surface tension, we repeated the
simulations in the bulk at 0.95 $T_m$, but with a small negative
hydrostatic pressure applied to the cell ($p\sim -1.8$ kbar, {\em
i.e.}, the same $p$ corresponding to the surface tension in the slab.)
This is equivalent to lowering the density of the liquid, bringing it
closer to that of the solid (see Fig.4).  As shown in Fig.3, the
slight decrease in liquid density in the bulk reproduces essentially
the observed increase of nucleation rates in the liquid slab.

\begin{figure}[b]
\begin{center}
\includegraphics[width=2.7in]{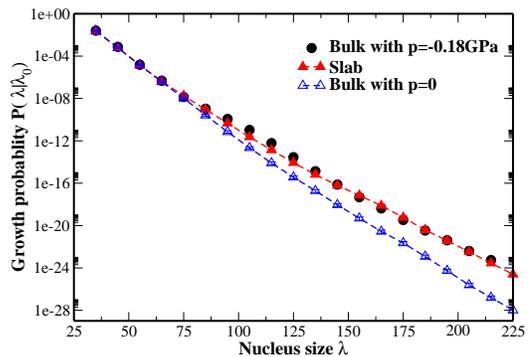}
\caption{(color online) Effect of $p$ on the homogeneous nucleation
  rate in the bulk liquid at $0.95T_m$. A small negative $p$, with the
  same magnitude as that induced by surface tension in the liquid
  slab, is applied on the bulk liquid. The resulting growth
  probability (black solid circles) almost coincides with that
  obtained for the liquid slab (red solid triangle).}
\label{f3}
\end{center}
\end{figure}

The calculated temperature dependent density change, as obtained for
the Tersoff Si, is shown in Fig.4 for both the liquid and the solid
near $T_m$. Note that at all $T$ in our simulation, the liquid slab is
under tension, as a result of the presence of surfaces. At 0.95 $T_m$,
liquid Si is about 1\% denser than the solid. Hence the formation of a
less dense solid Si nucleus is easier in the proximity of a
surface. At 0.79 $T_m$, the density of supercooled liquid falls below
that of the solid. In this case, nucleation at the surface involves a
higher energy barrier than in the bulk, and it is therefore not
preferred.  We also notice that the densities of diamond and liquid Si
become equal at about 0.86 $T_m$, where our simulations show no
difference in rates between surface and bulk nucleations. To further
elucidate the role of density, we also conducted simulations for Ge at
0.79 $T_m$ using the Tersoff potential, and for Si at 0.86 $T_m$
employing the Stillinger-Weber (SW) potential \cite{SW}. In both
cases, the liquids are denser than the solids, with 3\% and 7\%
density difference, for the Tersoff Ge and S-W Si,
respectively. Calculations show (see the insert of Fig. 1B) that
freezing in the slab is still preferred for both systems, consistent
with our analysis, and our results for Tersoff Si.

\begin{figure}[t]
\begin{center}
\includegraphics[width=2.7in]{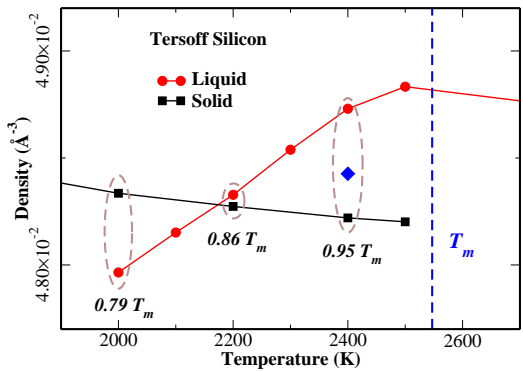}
\caption{Densities of liquid and diamond cubic silicon modeled by the
  Tersoff potential as functions of temperature near the melting point
  $T_m$. The blue diamond represents the density of liquid Si under a
  negative pressure P=-1.8 kbar at 0.95 $T_m$.}
\label{f4}
\end{center}
\end{figure}

While we found conditions under which crystallization is favored by
the presence of free surfaces, we emphasize that nucleation does not
occur exactly at the surface but instead in a subsurface region which
is a few atomic layers underneath. To understand that, we compute the
local pressure $p_{xx}(z)=\frac{1}{V(z)} \sum_{i\in V(z)}\left[m_i
v_{ix}^2+\frac{1}{2}\sum_{j\neq i, i\in V(z)}x_{ij}f_x(r_{ij})\right]$
and its correlation time $\tau_{xx}(z)=\int_0^{\infty}dt\frac{<\delta
p_{xx}(t,z)\delta
p_{xx}(0,z)>}{\sigma[p_{xx}(t,z)]\sigma[p_{xx}(0,z)]}$, where $\delta
p_{xx}(t,z)=p_{xx}(t,z)-<p_{xx}(t,z)>$,
$\sigma[p_{xx}(t,z)]=\sqrt{<p^2_{xx}(t,z)>-<p_{xx}(t,z)>^2}$ and the
bracket ``$<>$'' indicates ensemble averages, as a function of the
slab depth $z$. We find that the cluster tends to grow in the region
where the pressure field is non zero, however not where the pressure
field exhibits its minimum. This reflects the fact that while the
nucleation rate is dominated by free energy changes, the preferential
location of cluster formation is influenced by both static and
dynamical properties of the liquid. In fact, $\tau_{xx}(z)$, which
measures how rapidly pressure fluctuations decay, shows a maximum at
the surface, decaying with an oscillatory behavior towards the center
of the slab. Therefore, the preferential location of the nucleus is
the result of a subtle balance between the local static pressure and
its dynamical fluctuations. Details of $p_{xx}(z)$ and $\tau_{xx}(z)$
calculation will be given in a longer report.

Our calculations demonstrate that free surfaces, in addition to their
well-known role in initiating melting, can also be catalytic sites for
freezing in tetrahedral liquids with a negative slope, $dP/dT<0$, of
their melting lines. This unexpected result is related to the density
decrease occurring upon crystallization in these systems, and to
surface tension facilitating the initial nucleus formation. Our result
is consistent with recent experiment and simulations
\cite{Bhat,Molinero} showing that liquid Ge can be vitrified
(suppression of crystallization) by applying pressure. Our results
suggest that surface catalyzed nucleation should also be observed in
other tetrahedrally bonded materials showing density decrease upon
solidification. One interesting case is water, for which there is
experimental evidence suggesting surface crystallization in tiny water
droplets suspended in clouds \cite{Tabazadeh}. These findings can be
naturally explained by our results. We also notice that previous
direct MD simulations of water \cite{Vrbka} showed that ice
crystallizes in a subsurface region, consistent with our findings.

We thank D.C. Chrzan, A.F. Voter, and M. Parrinello for fruitful
discussions. This work was supported by DOE (contract numbers
DE-FG02-06ER46262 and DE-FC02-06ER25794).


\end{document}